\documentclass[journal]{vgtc}                      

\onlineid{0}

\vgtccategory{Research}

\title{\textbf{{\bonsai}}: A Mixed-Initiative Workspace for Human-AI\\Co-Development of Visual Analytics Applications} 
\cleantitle{BONSAI: A Mixed-Initiative Workspace for Human-AI Co-Development of Visual Analytics Applications}

\author{%
    \authororcid{Thilo Spinner}{0000-0002-1168-1804},
    \authororcid{Matthias Miller}{0000-0002-6281-2173},
    \authororcid{Fabian Sperrle-Roth}{0000-0002-2416-8639},
    \authororcid{Mennatallah El-Assady}{0000-0001-8526-2613}
}
\cleanauthor{Thilo Spinner, Matthias Miller, Fabian Sperrle-Roth, Mennatallah El-Assady}

\authorfooter{
    \item
    Thilo Spinner, Matthias Miller, Fabian Sperrle-Roth, and Mennatallah El-Assady are with ETH Zürich.
    E-mails: tspinner@ethz.ch, mmille@ethz.ch, fsperrle@ethz.ch, elassady@ethz.ch.
}

\usepackage{overpic}
\usepackage{wrapfig2}
\usepackage{tcolorbox}
\usepackage{tikz}
\usepackage{inlinegraphicx}
\usepackage{colortbl}
\usepackage{makecell}
\usepackage{xspace}

\newcommand{\bonsai}{\mbox{\textsc{bonsai}}\xspace}

\newcommand{\iviaparagraph}[1]{\refstepcounter{paragraph}\noindent\textbf{#1\ ---}\label{par:\theparagraph}}

\makeatletter
\newtcbox{\kwColorBoxSpecial}[1][]{on line,fontupper=\footnotesize\sffamily\bfseries\small,boxrule=0.5pt,arc=2pt,coltext=white,colback=#1,colframe=#1,boxsep=0pt,left=1.5pt,right=1.5pt,top=1.5pt,bottom=1.5pt}
\newcommand{\kwSpecial}[2]{%
    \begin{kwColorBoxSpecial}[#2]%
    {#1}%
    \end{kwColorBoxSpecial}%
    \xspace%
}

\makeatletter
\newcommand{\defEntity}[2]{%
    \expandafter\gdef\csname entity@color@#2\endcsname{#1}%
    \ifx
        \protect\@typeset@protect
        \kwSpecial{\phantomsection\label{entity:#2}#2}{#1}%
    \else
        #2%
    \fi
}
\newcommand{\refEntity}[1]{%
    \@ifundefined{entity@color@#1}{%
        \PackageError{config.tex}{%
            Entity `#1' referenced before definition%
        }{%
            Use \string\defEntity\{<color>\}\{#1\} before calling
            \string\refEntity\{#1\}.%
        }%
    }{%
        \ifx
            \protect\@typeset@protect
            \hyperref[entity:#1]{\kwSpecial{#1}{\csname entity@color@#1\endcsname}}%
        \else
            #1%
        \fi
    }%
}
\makeatother

\definecolor{PhaseAColor}{HTML}{305bab}
\definecolor{PhaseBColor}{HTML}{067429}
\definecolor{PhaseCColor}{HTML}{6631d7}
\definecolor{PhaseDColor}{HTML}{af7e04}

\definecolor{ChellengeColor}{HTML}{baa586}
\definecolor{RequirementColor}{HTML}{9b9ba7}

\abstract{%
    Developing Visual Analytics (VA) applications requires integrating complex machine learning models with expressive interactive interfaces.
    Developers face a stark trade-off: building tightly-coupled monoliths plagued by fragile interdependencies, or relying on restrictive, simplistic frameworks.
    Meanwhile, unconstrained, single-shot AI code generation promises speed but yields unstructured, unauditable chaos.
    The core challenge is combining the control and expressiveness of custom development with the efficiency of AI generation under strict constraints.
    To address this, we introduce \bonsai, a mixed-initiative workspace for the multi-agent co-development of VA applications.
    \bonsai utilizes a modular four-layer architecture (hardware, services, orchestration, application) that allows human and AI developers to independently contribute reusable components.
    The workspace incorporates this architecture into a structured four-phase development process (plan, design, monitor, and review), ensuring distributed agency and full provenance, where all human and AI contributions are structurally bounded and tracked. We evaluate \bonsai through case studies demonstrating the efficient creation of novel tools and the rapid reconstruction of complex VA applications directly from research paper descriptions. Ultimately, this paper contributes a conceptual workflow, a scalable architecture, and an integrated system that successfully balances AI's generative speed with the structural rigor required for complex VA development.
}

\teaser{
    \centering
    \includegraphics[width=1\linewidth]{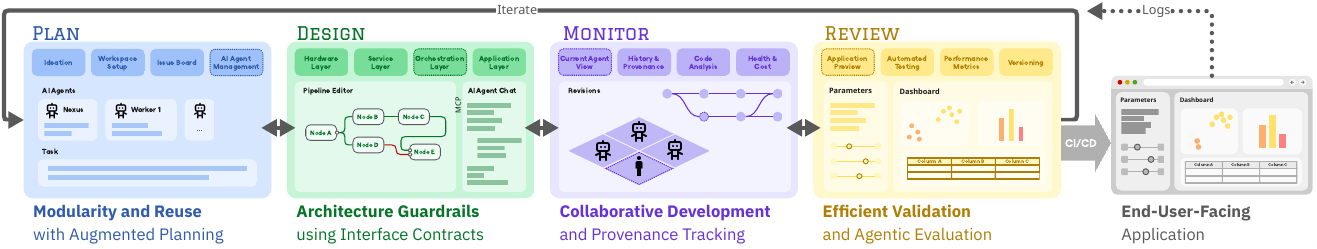}
    \vskip -0.1em
    \caption{%
        The structured four-phase workflow of the \bonsai mixed-initiative workspace.
        The human-driven process begins with (1) \textbf{Plan} to ensure modularity and reuse through augmented task and agent management.
        In the (2) \textbf{Design} phase, human and AI agents collaborate within bounded contexts, enforcing architectural guardrails via strict interface contracts across the four system layers.
        The (3) \textbf{Monitor} phase enables safe collaborative development through comprehensive provenance tracking of all multi-agent interactions.
        The (4) \textbf{Review} phase facilitates efficient validation of the resulting Visual Analytics application through agentic evaluation, automated testing, and performance dashboards.
        Finally, the application is deployed to be accessed by the end-user.
    }
    \vskip -0.1em
    \label{fig:tasks}
}

\graphicspath{{figs/}{figures/}{pictures/}{images/}{./}} 

\usepackage{tabu}                      
\usepackage{booktabs}                  
\usepackage{lipsum}                    
\usepackage{mwe}                       
\usepackage{ccicons}                   

\usepackage{mathptmx}                  
\usepackage{microtype}
\begin{document}


\firstsection{Introduction}
\maketitle

Developing data-driven Visual Analytics (VA) applications is inherently complex, requiring tight integration of machine learning models, intensive data-processing pipelines, expressive interaction design, and heterogeneous execution environments~\cite{keim2008va}.
Developers face persistent challenges in decomposition and reuse across projects, such as extracting a novel visualization technique from a prototype, sharing a reusable view, or reconstructing a complex system described in a research paper.
While these demands do not strictly forbid monolithic implementations, tightly coupled architectures silently tangle concerns across the resulting systems, making it difficult to reason about, audit, and extend~\cite{parnas1972decomposition}.

Recently, agent-assisted and fully agentic coding workflows have matured significantly~\cite{yang2024sweagent}.
Current best practices---leveraging stronger frontier models, structured plan files, agent skill constraints, and automated review steps---can dramatically improve development throughput and catch shallow failures.
However, this rapid pace introduces a new sociotechnical challenge often colloquially termed ``vibe coding''~\cite{vibecoding2025empirical}: a momentum-driven development style where the sheer speed and ease of generation motivate developers to rely on intuition rather than critical oversight.
While frontier tools allow developers to read along and steer the process, they rarely \textit{enforce} active comprehension.
Consequently, developers are easily incentivized to disengage from the causal story of how outputs are produced, leading to misplaced confidence in system layers where the human reviewer no longer possesses a stable mental model~\cite{barke2023grounded,perry2023insecure,pearce2022asleep}.
The resulting gap is a severe lack of fine-grained provenance: there is no legible record of why an integration decision survived review, or how human and machine edits intertwined~\cite{souza2025provagent,agenttrace2026}.

To restore intelligibility and control, the underlying architecture must enforce a strict separation of concerns.
Modularity and explicit interface contracts are standard software-engineering responses to exactly this kind of boundary pressure.
We adopt a layered core architecture (hardware, services, orchestration, application), not to claim novelty for layering as such, but as a disciplined, machine-enforceable scaffold.

Interface contracts and modularity operate both \textit{within} and \textit{across} these layers.
Vertically, each layer enforces its own standardization: the hardware layer achieves consistency across heterogeneous platforms through Kubernetes, abstracting bare-metal and cloud nodes behind a uniform scheduling interface; the service layer exposes independently deployable microservices with typed, versioned interfaces; and the orchestration layer provides centralized management components (authentication, service registry, workflow orchestrator) powering DAG-based workflows that enable design-time type checking and selective recomputation.
Horizontally, each layer exposes only a minimal typed surface to the layer above: context distillation optimizes the information scope that any developer or agent needs to hold at once, and the abstraction barrier prevents upper layers from reaching through to bypass or alter lower-layer implementations.

By applying these standard good practices, we transform the unconstrained ``\textit{AI-does-everything}'' sandbox into targeted, manageable tasks.
An AI agent is bound to a specific layer and constrained by its interface contract, which prevents context-window saturation~\cite{liu2024lostinthemiddle} and keeps responsibilities and review surfaces explicit---a prerequisite for legible human--AI co-development.
Furthermore, it enables VA developers to focus exclusively on specific layers according to their expertise---for example, application developers can concentrate on UI/UX and workflow logic, service developers on atomic data-processing algorithms, and DevOps engineers on the underlying execution environments.
Under this level of bounded expert control, agentic systems unlock massive potential, including the rapid reconstruction of complex VA applications directly from research papers that lack published source code.

To operationalize this scaffold, we introduce \textbf{\bonsai}, a mixed-initiative workspace for the human--AI co-development of data-driven VA applications.
Its name signals our stance: like cultivating a \bonsai tree, generative code changes represent organic growth that still demands deliberate structure, pruning, and guided oversight if the product is to stay robust and inspectable.
The workspace organizes collaboration through four phases---Plan, Design, Monitor, and Review.
This structured workflow ensures that oversight, provenance tracking, and a balanced distribution of agency remain visible and actionable throughout development, rather than collapsing into a single opaque transcript.

Specifically, this paper makes four principal contributions:
(1)~a conceptual analysis of human--AI co-development tailored to VA applications, emphasizing the necessity of modularity, reuse, guardrails, interface contracts, provenance tracking, and agency distribution;
(2)~a layered core architecture encompassing four distinct levels (hardware, services, orchestration, and application) that provides the structural foundation for safe multi-agent interaction;
(3)~the introduction of the \bonsai workspace, an integrated system for human--AI collaboration that operationalizes this architecture through a four-phase development process;
and (4)~validation of the approach via use cases demonstrating the system's effectiveness in rapidly reconstructing complex VA applications from research paper descriptions.

\section{Related Work}

\bonsai sits at the intersection of workflow orchestration, visualization architecture design, mixed-initiative human-AI collaboration, provenance tracking, and agentic software engineering.
We structure this section along these dimensions, highlighting how existing approaches address parts of the problem but leave gaps that \bonsai fills.

\iviaparagraph{Workflow Management and Data Pipeline Tools} The orchestration of multi-step computational workflows has a long history in data engineering.
Apache Airflow~\cite{airflow2024} popularized workflows as directed acyclic graphs (DAGs) in Python, providing scheduling, dependency resolution, and monitoring.
More recent platforms such as Prefect~\cite{prefect2024} and Kestra~\cite{kestra2024} decouple orchestration from business logic, with Kestra using declarative YAML definitions for language-agnostic execution.
Dagster~\cite{dagster2024} introduces \emph{software-defined assets}, treating data lineage as a first-class citizen.
This principle resonates with \bonsai's typed interface contracts.
KNIME~\cite{berthold2009knime} offers a visual node-based interface for composing analytical pipelines.
However, these tools target batch processing and do not address the interactive feedback loops of VA applications nor bounded human-AI co-development. \bonsai extends their declarative philosophy with a layered architecture for VA pipeline construction, where services, orchestration logic, and front-ends co-evolve under human oversight.

\iviaparagraph{Visualization Architectures and Design Constraints}
Satyanarayan et al.\ introduced Vega-Lite~\cite{satyanarayan2017vegalite} as a high-level grammar of interactive graphics on Vega's reactive dataflow~\cite{satyanarayan2016reactivevega}.
Moritz et al.\ formalized visualization design as a constraint satisfaction problem in Draco~\cite{moritz2019draco}, enabling automated encoding recommendations.
Heer and Moritz proposed Mosaic~\cite{heer2024mosaic}, decoupling visualization clients from a scalable query engine---a separation of concerns paralleling \bonsai's layered design.
Wongsuphasawat et al.\ showed with Voyager~2~\cite{wongsuphasawat2017voyager2} how partial specifications enable mixed-initiative exploration of design spaces, and Ding et al.~\cite{ding2025guardrails} found that shared visual structures for human--AI hypothesis exploration act as ``guardrails'' focusing thinking without constraining creativity---validating \bonsai's pipeline editors as cognitive scaffolds.
These systems advance visualization rendering, recommendation, and scalable querying, but target the \emph{use} of visualizations, not the \emph{co-development} of full VA applications including backend data processing pipelines.

\iviaparagraph{Mixed-Initiative Visual Analytics and Guidance}
Mixed-initiative interaction---where humans and agents dynamically share control---was formalized by Horvitz~\cite{horvitz1999mixed} in his foundational principles for mixed-initiative user interfaces.
Amershi et al.~\cite{amershi2019guidelines} later consolidated practical guidelines for human--AI interaction that have become a reference in the field.
Ceneda et al.~\cite{ceneda2017guidance} characterized guidance as closing knowledge gaps and proposed a designer model~\cite{ceneda2020guide} for the what, when, and how of guidance.
Sperrle et al.\ introduced co-adaptive guidance for learning user preferences~\cite{sperrle2021learning,sperrle2021coadaptive} and contributed Lotse~\cite{sperrle2023lotse}, a library for deploying guidance strategies in running VA systems.
El-Assady et al.~\cite{elassady2019speculative} proposed speculative execution for VA, enabling proactive computation of competing model states.
St{\"a}hle et al.~\cite{stahle2025designspace} presented a design space for agents in VA, while Monadjemi et al.~\cite{monadjemi2025scopingreview} confirmed the breadth of mixed-initiative VA research while identifying the lack of frameworks moving beyond analysis assistance toward development support.
Across this work, mixed-initiative interaction targets \emph{data analysis}. \bonsai transfers these principles to the \emph{development} process, guiding developers and agents through the co-construction of VA applications and distributing agency across architectural layers and phases.

\iviaparagraph{Provenance Tracking in Visual Analytics}
Provenance---the systematic recording of history and lineage of analytical artifacts---has been extensively studied in visualization.
Ragan et al.~\cite{ragan2016provenance} proposed a taxonomy distinguishing data, visualization, interaction, insight, and rationale provenance.
Xu et al.~\cite{xu2020provenance} provided a comprehensive survey establishing a unified framework for provenance in VA.
On the systems side, VisTrails~\cite{callahan2006vistrails} represents workflow evolution as a version tree, and AVOCADO~\cite{stitz2016avocado} manages dense provenance graphs through hierarchical aggregation---a challenge mirrored in \bonsai's multi-agent development histories.
CLUE~\cite{gratzl2016clue} enables capturing and explaining provenance with branching, Trrack~\cite{cutler2020trrack} provides a reusable tracking library, and Loops~\cite{eckelt2024loops} extends provenance into computational notebooks.
In agent provenance, PROV-AGENT~\cite{souza2025provagent} extends W3C PROV with agent-centric entities, while Cursor's Agent Trace~\cite{agenttrace2026} addresses AI code attribution at the file level.
These contributions provide mature models for tracking user interactions and agent actions, but none address provenance in VA \emph{development}---tracking who contributed what, at which layer, and during which phase. \bonsai treats agency attribution as a first-class dimension: every modification across all four layers is recorded with its actor, enabling developers to audit, compare, and roll back human and AI contributions throughout the development lifecycle.

\iviaparagraph{Agentic Software Engineering}
AI-assisted coding tools---GitHub Copilot, Cursor~\cite{cursor2025}, Windsurf~\cite{windsurf2026}, JetBrains AI~\cite{jetbrainsjunie2025}, Claude Code~\cite{claudecode2025}, and OpenAI's Codex~\cite{codex2025}---integrate large language models into development environments for code generation and increasingly autonomous task execution.
In the multi-agent space, MetaGPT~\cite{hong2024metagpt} encodes Standardized Operating Procedures into multi-agent pipelines, AutoGen~\cite{wu2023autogen} orchestrates configurable multi-agent conversations, and Magentic-One~\cite{fourney2024magneticone} employs a dual-loop orchestrator.
On the protocol level, Anthropic's Model Context Protocol~\cite{mcp2024} standardizes agent-to-tool communication, Google's Agent2Agent Protocol~\cite{a2a2025} addresses inter-agent discovery, and the LLM Delegate Protocol~\cite{prakash2026ldp} adds identity-aware routing and structured provenance---notably showing that self-reported quality scores degrade routing below random baselines.
Empirically, Peng et al.~\cite{peng2023copilotproductivity} found significant productivity gains, while Pearce et al.~\cite{pearce2022asleep} and Perry et al.~\cite{perry2023insecure} showed that AI-generated code frequently contains vulnerabilities and that developers overestimate its security.
These findings expose three gaps:
(1)~existing tools operate as monolithic agents without distribution of agency across architectural boundaries;
(2)~provenance is limited to git-level attribution, missing fine-grained human--AI interplay; and
(3)~the absence of guardrails---typed contracts, bounded contexts---leads to context-window saturation and opaque generation as projects grow.
\bonsai addresses all three through bounded, layer-specific AI Development Units, typed interface contracts, and comprehensive agency provenance across a four-phase development process.

\section{Problem Characterization}
To transition from the limitations of monolithic development to a robust, AI-assisted paradigm, we must formalize the system's boundaries.

\iviaparagraph{Formative Methodology}
To ground our design rationales and architectural decisions in practice, we used an iterative, human-centered process over three years: we first developed and validated the layered architecture \textit{without AI} in multiple higher-education courses (200+ students per cohort) building complex VA pipelines, then used longitudinal observations of recurrent failures (e.g., dependency conflicts, brittle cross-layer contracts, and poor reproducibility) to design the \bonsai workspace and scope generative-AI support as guardrails, addressing real development bottlenecks rather than hypothetical edge cases.

\iviaparagraph{Target User Groups}
Standard VA development often assumes a homogeneous ``developer'' role, but real systems involve distinct roles.
To integrate AI agents safely, \bonsai aligns its abstractions with four groups: \textbf{End-Users} audit and steer application behavior via feedback, \textbf{Application Developers} build UI and interaction logic, \textbf{Service Developers} build algorithms, models, and service contracts, and \textbf{DevOps-Engineers} take care of deployment, orchestration, and security.
This separation lets both humans and AI assistants operate with the right layer-specific context, avoiding cognitive and computational overload.

\subsection{Challenges}

First, developers face \textbf{\defEntity{ChellengeColor}{ch01} \textit{Tightly-Coupled Monoliths and Dependency Hell}}.
Traditionally, VA tools are built as highly specialized, one-off monolithic architectures, leading to tremendous dependency trees virtually impossible to maintain.
Updating individual packages often entails an avalanche of incompatibilities between upstream dependencies which are hard or even impossible to resolve.

Related, accessibility of useful software components is limited by the \textbf{\defEntity{ChellengeColor}{ch02} \textit{(Re-)Usability Bottleneck}}.
Extracting useful standalone functionalities from a monolith is non-trivial: components are typically entangled with application-specific logic and glue code, making isolation labor-intensive and error-prone.
The problem is compounded for research-driven VA tools, where novel techniques are frequently published without well-maintained implementations, forcing developers to spend significant effort re-implementing algorithms from paper descriptions before any integration work can begin.

Second, as generative AI is introduced to speed up development, systems suffer from \textbf{\defEntity{ChellengeColor}{ch03}\textit{Architectural Drift and Context Degradation}}.
While modern agentic harnesses are increasingly capable, the quality of AI-generated code is ultimately bounded by the standards the project itself establishes: a codebase with clear boundaries, strict conventions, and well-defined interfaces naturally guides the AI toward coherent, maintainable outputs.
Unstructured or entangled codebases instead cause the AI to replicate and amplify existing flaws.
This accelerates technical debt rather than alleviating it.
Without bounded layers, the AI’s context is saturated with suboptimal patterns and details rather than the targeted interface contracts needed for reliable generation.

The rapid pace of AI-driven development exacerbates \textbf{\defEntity{ChellengeColor}{ch04}\textit{Loss of Semantic Provenance in Co-Creation}}.
While AI harnesses are now proficient at utilizing feature branches and committing code regularly, standard version control diffs fail to capture \textit{intent}:
\texttt{git} histories do not always record \textit{why} an AI made a specific integration decision, what prompted the generation, or how agency was distributed between the human and the machine.
Without this semantic provenance, developers lose the ability to meaningfully audit or steer the system's evolution.

Finally, VA applications deployed on heterogeneous, multi-tenant platforms face \textbf{\defEntity{ChellengeColor}{ch05}\textit{Fragile Compliance Across Heterogeneous Deployment Stacks}}.
Applications routinely carry operational and regulatory requirements: data must not leave a particular jurisdiction, certain computations require GPU-enabled hardware, sensitive payloads must not be routed through external services.
While these requirements originate at the application level, their enforcement must happen at the infrastructure level.
Without a structural bridge between the two, compliance depends on manual configuration and organizational convention at every layer independently: a developer who specifies that clinical data must remain on-premise has no guarantee that an external API will respect this.
As the number of services, workflows, and execution environments grows, this ad-hoc approach becomes increasingly fragile and difficult to audit.

\subsection{System Requirements}
From these challenges, we derive three key requirements for safe, multi-agent co-development in the \bonsai workspace.

\iviaparagraph{\defEntity{RequirementColor}{R1}Modularity and Reuse}
To resolve \refEntity{ch01} and \refEntity{ch02}, the system must enforce a strict separation of concerns, breaking the paradigm of isolated, single-use prototypes.
Functionalities must be encapsulated into atomic, operationally independent microservices.
By extracting these building blocks and preventing deep inter-dependencies, the architecture must guarantee that newly developed analytical techniques and interface components are natively reusable across multiple projects and easily deployable by different teams, rather than remaining locked away in custom glue code.

\iviaparagraph{\defEntity{RequirementColor}{R2}Guardrails and Context Distillation}
To address \refEntity{ch03} and reliably manage \refEntity{ch05}, the architecture must provide strict boundaries that naturally distill the context for both human developers and agentic AI.
Rather than exposing an unstructured full-stack codebase, the system must establish clear, formal interface definitions (e.g., semantic API documentation and API contracts) that enable design-time dependency resolution and validity checks.
This deliberate context distillation ensures that any given agent's context is rich enough to complete a specific task, yet entirely pruned of redundant, irrelevant, or cross-layer implementation details.
By constraining the action space to these well-documented, localized boundaries, generated outputs remain small and highly traceable.
It is precisely this reduction in cognitive load and complexity that allows human developers to meaningfully review both their own and the AI's contributions, guaranteeing they remain in full, confident control over the development process.

\iviaparagraph{\defEntity{RequirementColor}{R3}Provenance Tracking and Agency Sharing}
To overcome \refEntity{ch04}, the system must explicitly reject unconstrained generative workflows in favor of formal agency sharing, where humans and AI collaborate across well-defined boundaries.
In a mixed-initiative environment, robust accountability requires semantic provenance: tracking not just \textit{what} code changed, but \textit{who} (human or AI) authored it, and \textit{why}.
Crucially, this provenance tracking cannot rely solely on generic, one-size-fits-all version control logs.
It must be tailored to each layer and treat agency handoff points as first-class citizens in the system's state history, enabling developers to audit, steer, or roll back the co-creation process at any time.

\section{Bonsai's Layered Core}
\label{sec:architecture}

To satisfy the aforementioned requirements (\refEntity{R1}, \refEntity{R2}, \refEntity{R3}) and physically isolate the operational domains of our target user groups, we introduce a four-layered architecture.
Rather than relying on soft coding conventions, this infrastructure structurally enforces a strict separation of concerns.
By untangling the complex dependencies of traditional VA pipelines into discrete, manageable environments, the architecture natively bounds the context for both human developers and their assistive AI agents.
As detailed in the following subsections, each layer contributes distinct and complementary mechanisms that together operationalize all three requirements:
modularity and reuse (\refEntity{R1}) through strict service encapsulation and a shared registry that makes components immediately composable across projects;
guardrails and context distillation (\refEntity{R2}) embedded structurally at every level rather than imposed by post-hoc convention;
and the architectural foundations for provenance and agency attribution (\refEntity{R3}), which the \bonsai workspace (see \cref{sec:bonsai-workspace}) operationalizes fully through its four development phases.

\begin{figure}
    \centering
    \includegraphics[width=\linewidth]{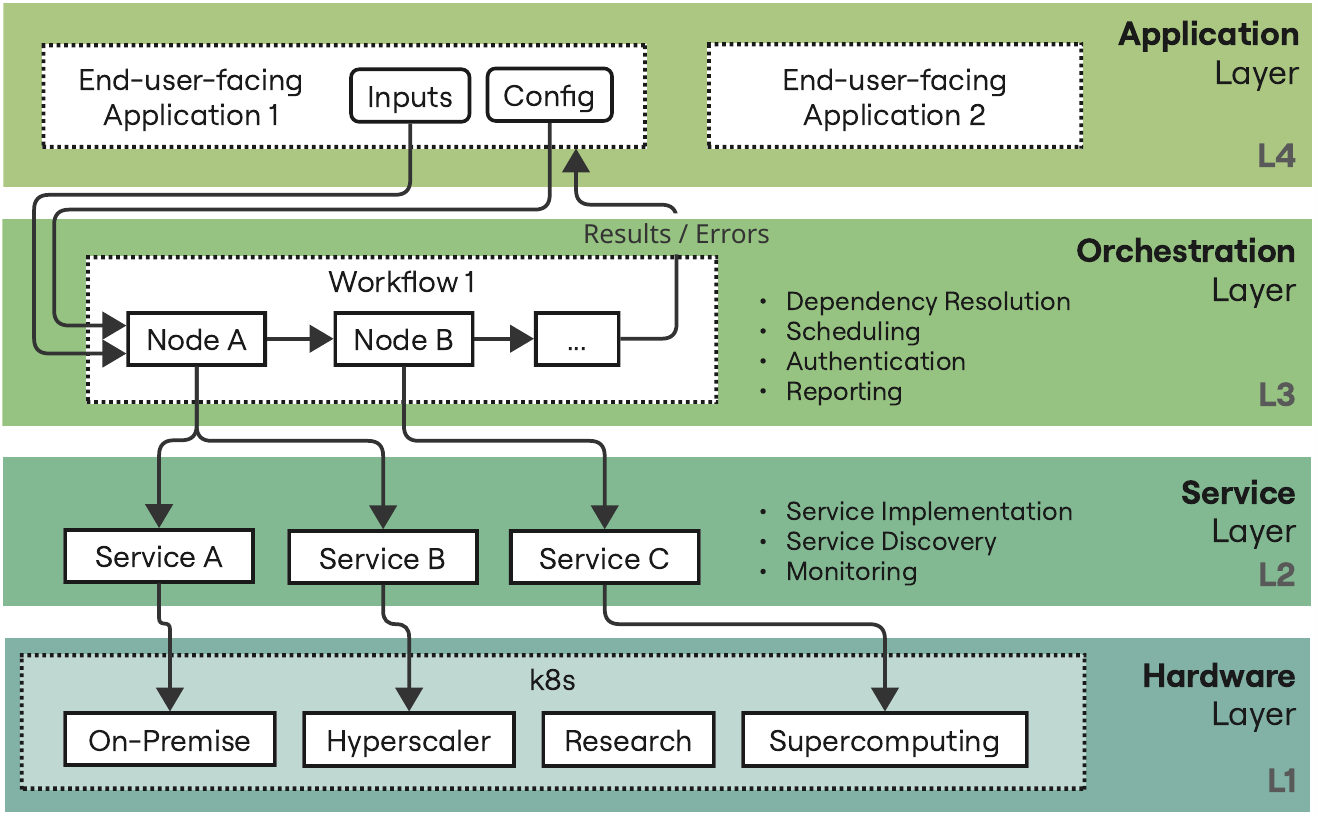}
    \vspace{-1.5em}
    \caption{The layered core architecture. The Application Layer provides a low-code environment for rapid pipeline configuration. The Orchestration Layer manages workflow execution and distributes tasks to the Service Layer, where independent microservices process data sequentially in accordance with strict API specifications. The underlying Hardware Layer abstracts diverse, heterogeneous resources via Kubernetes.}
    \label{fig:architecture}
    \vspace{-1em}
\end{figure}

\subsection{Core Architecture}
\label{sec:core architecture}
As illustrated in \cref{fig:architecture}, the middleware is organized into four hierarchical layers, each abstracting complexity from the one below.
At the base, the \emph{Hardware Layer} (L1) exposes heterogeneous compute resources---on-premise clusters, hyperscalers, research, and supercomputing---through a unified Kubernetes abstraction, enabling policy-driven scheduling that transparently matches workload requirements to compliant execution environments.
The \emph{Service Layer} (L2) hosts independently deployable microservices, each encapsulating a single AI or data-processing capability behind a typed OpenAPI contract.
The \emph{Orchestration Layer} (L3) maintains a shared service registry that handles discovery and monitoring of all admitted L2 services; it accepts workflow definitions as directed acyclic graphs and coordinates their execution: resolving inter-node dependencies, scheduling service calls, enforcing authentication, and routing results---or errors---back to the caller.
The \emph{Application Layer} (L4) exposes this machinery to end users as a low-code environment in which applications compose workflows from registry services, supply inputs and configuration parameters, and consume structured results interactively.

\subsubsection*{\textbf{L1: Hardware Layer \& Policy-Driven Scheduling}}
The Hardware Layer unifies disparate execution environments, such as on-premise clusters, HPC systems, or cloud providers, 
into a single logical resource pool managed via Kubernetes, providing the infrastructure required for heterogeneous VA execution.

Beyond resource pooling, it implements a three-part cross-layer constraint satisfaction mechanism that directly addresses \refEntity{R2}: application-level requirements are reliably enforced throughout the entire execution stack, not merely declared at the top.
To this end, first, all entities in the stack (hardware environments, services, workflows, and applications) are \textbf{tagged} with a set of descriptive labels encoding properties along multiple dimensions, including data jurisdiction, confidentiality level, runtime behavior (e.g., absence of external network dependencies), hardware capabilities, and regulatory compliance.
Second, applications and workflows explicitly \textbf{declare} the constraints they require the layers below to satisfy; these declarations are versioned alongside the workflow definition, making compliance requirements a reproducible part of the workflow artifact rather than an external, mutable configuration.
Third, at scheduling time, the platform \textbf{matches} constraints through the stack.
Critically, a constraint that cannot be satisfied blocks execution and surfaces a diagnostic error, rather than silently falling back to a non-compliant environment.
For example, if a medical application is tagged as \textsc{HIPAA-compliant}, the orchestration layer will ensure its workloads call only compliant services.
The hardware scheduler will ensure execution occurs exclusively on secure, on-premise nodes without requiring manual intervention per workflow.

\subsubsection*{\textbf{L2: Service Layer \& Service Development}}
The Service Layer is the domain of Service Developers and specialized backend AI agents.
A \emph{(micro)service} is a small, independently deployable software component that encapsulates a single AI or data-processing capability and exposes it through a typed, validated OpenAPI contract.
The microservice-based architecture is strictly motivated at this layer to address \refEntity{R1}: by enforcing strong encapsulation, services remain independently maintainable, can scale through replication, and are reusable across disparate VA projects without introducing tangled cross-project dependencies.

\iviaparagraph{Admission Gate} In L2, developers build and deploy atomic data-processing algorithms and machine learning models.
To be admitted to the Service Registry and, thus, become discoverable by the orchestration layer and eligible for hardware scheduling, every service must satisfy a set of mandatory interface requirements: a RESTful API, a strongly typed and complete OpenAPI specification, a health endpoint for runtime monitoring, semantic versioning, and a commitment to backward compatibility within a given endpoint version.
These admission checks are enforced before a service becomes visible to any workflow or AI agent, ensuring that only well-specified, compatible components enter the shared ecosystem.

\iviaparagraph{Developer and Agent Guidance} To guide both human developers and AI coding agents in meeting these requirements, \bonsai provides skill files (curated, machine-readable documents encoding coding conventions, architectural patterns, and layer-specific guardrails) alongside standardized scaffolds that serve as fully conformant service templates.
Together, these resources operationalize \refEntity{R2} at the service level: rather than exposing agents to an unstructured codebase, the OpenAPI contract provides a semantically complete, implementation-agnostic description of each service's capabilities---precisely scoping what any developer or agent needs to know to integrate or extend a component, without access to irrelevant internal detail.
For more details on how AI agents implement services as part of their workflow, see \cref{sec:integration-of-agentic-ai}.

\subsubsection*{\textbf{L3: Orchestration Layer \& Control Plane}}
The Orchestration Layer acts as the central control plane, bridging services (L2) with applications (L4).
It manages real-time execution, service discovery, and automatic dependency resolution, while serving as the architectural locus for three complementary guardrail mechanisms.

\iviaparagraph{Centralized Identity and Access Management} L3 provides a stateless authentication gateway that intercepts every inbound service call, validates the caller's token against the Service Registry, and enforces uniform access control across the entire ecosystem.
Services are required to defer all authentication and authorization to this gateway and must not implement their own identity and access management (IAM) logic.
This constraint eliminates security vulnerabilities arising from divergent implementations, prevents inconsistent access patterns, and ensures that access control is a uniform, centrally auditable property of the platform, providing the actor-level attribution log that is a structural prerequisite for \refEntity{R3}.

\iviaparagraph{Structured Workflow Composition} Application logic in L3 is expressed as directed acyclic graphs (DAGs) through a workflow designer that enforces a strict visual and textual grammar: nodes represent registered L2 services or platform-provided control-flow constructs (e.g., conditional branches, parallel execution blocks), and edges represent typed data flows.
Compatibility between connected nodes is validated at design time by the \emph{CType} structural type system---where a CType is a named, structured data type composed of typed fields, each of which is either a primitive type (integer, string, boolean, etc.) or another CType---which checks that the output types of upstream nodes precisely match the expected input types of downstream nodes, catching structural mismatches already during design-time.
Once validated, the DAG is automatically transpiled into an executable flow and dispatched to the orchestration engine, which schedules service calls in dependency order.
When an end-user or agent adjusts a parameter mid-workflow, the dependency graph is resolved, and only the affected downstream nodes are re-executed—minimizing redundant computation and ensuring responsive UI interactions.
Because service discovery is continuous, newly registered L2 services immediately become composable nodes without requiring platform downtime or reconfiguration, leveraging the requirements and admission control mechanisms implemented by the service layer.
This composability from a shared, versioned registry directly operationalizes \refEntity{R1}: a service developed once for any project becomes immediately available for reuse across all workflows without modification or duplication.

\subsubsection*{\textbf{L4: Application Layer \& User Interaction}}
The Application Layer is the top-most abstraction, where Application Developers implement the End-User-facing VA application.
\iviaparagraph{Implementation Openness} Unlike the lower layers, L4 intentionally provides the greatest degree of freedom: because VA applications are highly individual in their visual languages, interaction paradigms, and domain-specific requirements, no single framework or rigid template could accommodate the full range of expressiveness needed.
Front-end developers may therefore use any framework of their choice to build custom, stateful user interfaces tailored to their analytical context.

To guide both human developers and UI-focused AI coding agents through this open-ended layer, \bonsai provides skill files, component templates, and curated library recommendations that encode best practices for connecting applications to the underlying platform~\cite{Gyarmati2025StructuredVisualizationDesign}.
These resources reduce decision overhead without prescribing a fixed implementation, operationalizing \refEntity{R2} at the application level within the inherent constraints of a free-form layer.

\iviaparagraph{Platform Integration} The application communicates with L3 via well-defined REST endpoints to discover workflow runs, fetch state histories, retrieve analytical results, and trigger selective re-executions.
When an end-user adjusts a parameter, the UI must push the change to the orchestration layer; L3 resolves the DAG, re-executes only the affected downstream nodes, and returns the updated state.
This clean separation allows Application Developers (and UI-focused AI agents) to focus entirely on visual components, UX workflow logic, and stateful interaction without managing service internals or infrastructure scaling.

Importantly, the only formal guarantees the architecture can provide at L4 are the typed input and output contracts of the L3 pipelines and L2 services the application consumes.
Beyond these contracts, correctness, visual quality, and interaction design remain the responsibility of the Application Developer---a deliberate boundary that preserves the expressiveness required for domain-specific VA tools while ensuring that the integration surface remains well-defined and auditable.

\subsection{Implementation Details}

The following paragraphs describe the key mechanisms that realize the architectural principles outlined above.

\iviaparagraph{Service Discovery}
Services deployed to designated Kubernetes namespaces are automatically discovered by the Service Registry.
Admission controllers verify that each deployment satisfies the platform's requirements before registration proceeds.
Services hosted outside the cluster can alternatively be registered manually.
In both cases, the OpenAPI documentation of the service is checked against a validation endpoint, ensuring the API's structure, its endpoints, and its typings match the specified quality criteria.
If the validation succeeds, a newly added service enters a \emph{Pending~Review} state visible to platform administrators.
Once reviewed, one user (usually the developer) is appointed as the service's manager and is responsible for governing access.

\iviaparagraph{Authorization and Authentication}
The Service Registry enforces fine-grained access control over service resources through OAuth 2.0 Bearer tokens\cite{Jones2012Oauth20Authorizationa}.
The ingress controller validates the Authorization header\cite{Nielsen1996HypertextTransferProtocol} of every incoming service call against a stateless authorization gateway, which in turn validates the token against the access control endpoints of the Service Registry.
Access to sensitive information held by a service is granted by possessing the resource identifier itself, eliminating the need for additional per-resource permission checks (but requiring that the resource identifier be treated as a secret).

\iviaparagraph{Workflow Composition, Dispatch, and I/O}
Workflows are authored either through a visual canvas editor or by directly editing YAML, with both representations kept in sync.
The editor exposes registered services as typed, drag-and-drop nodes whose connectable handles and static parameter panels are derived automatically from the service's OpenAPI schema: fields annotated with \texttt{x-parameter:\;true} appear as configuration knobs.
In contrast, all other request-body fields become wired inputs that can receive data from upstream nodes.
The canonical workflow artifact is a YAML revision stored in the portal's database; the underlying orchestration engine Kestra~\cite{kestra2024} is never given a persistent copy.
Instead, each execution follows an ephemeral dispatch pattern: the portal injects an execution-scoped identifier into the YAML, registers the flow with Kestra's REST API, starts an execution, and streams the resulting Server-Sent Event (SSE) progress feed.
Once a terminal state is reached, logs and any file-typed outputs are copied to the platform's object store, and the Kestra flow is deleted.
Scalar inputs are passed as typed key-value pairs (string, int, float, boolean, json).
File inputs are resolved from the platform's object store before submission.
For very large binary objects, it is recommended to instead use a Kestra plugin node that directly retrieves the file from a storage backend (e.g., S3).
Applications trigger workflows through a webhook or through the platform's MCP interface, supplying input values and optional file references.
Callers poll the execution state endpoint and retrieve output artifacts via presigned object-store URLs once execution succeeds.

\section{The \bonsai Workspace}
\label{sec:bonsai-workspace}

\begin{figure}
    \centering
    \includegraphics[width=\linewidth]{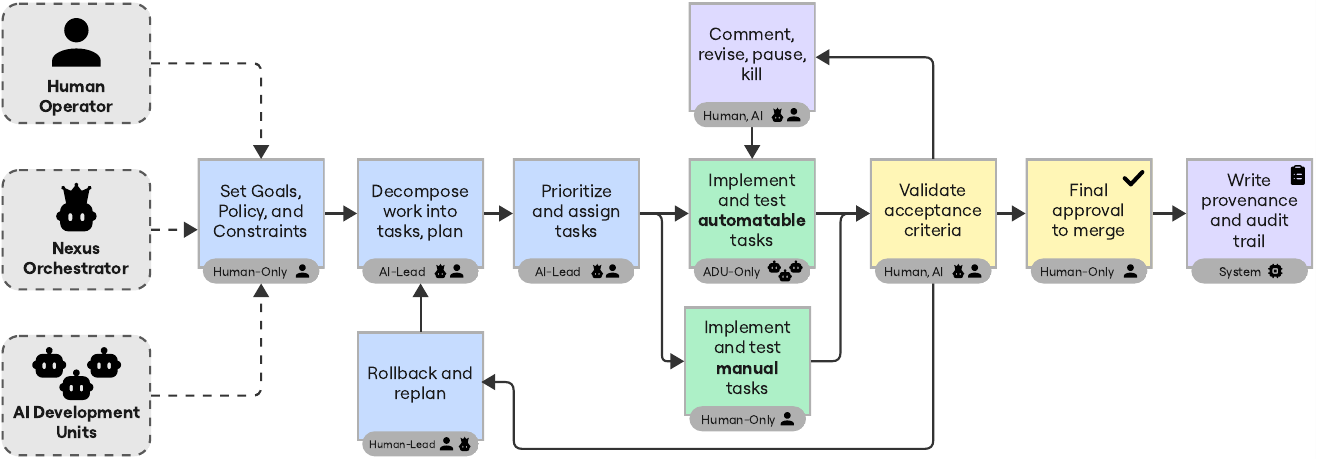}
    \caption{An exemplary human-AI collaborative Bonsai workflow, illustrating a natural distribution of agency across phases.%
    Human-Only, AI-Lead, and mixed Human-AI phases alternate to balance automation with oversight, with feedback loops enabling iterative revision and rollback throughout.}
    \label{fig:human-ai-collaborative-workflow}
    \vspace{-1em}
\end{figure}

The layered core provides structural foundation and interface contracts, but infrastructure alone is insufficient for successful human-ai collaboration: the process must keep the human in the loop and prevent the AI from acting as an opaque black box.
We therefore introduce the \bonsai Workspace, a mixed-initiative environment that operationalizes the architecture through a continuous, four-phase workflow: Plan, Design, Monitor, and Review~(see \cref{fig:tasks}).
These phases make no assumption about agency distribution---either party could, in principle, run all four alone.
In practice, a natural division emerges (\cref{fig:human-ai-collaborative-workflow}): humans set goals, policies, and constraints, while the top-level AI orchestrator (Nexus) decomposes tasks and delegates them via mid-level coordinators (Squad Leads, SLs) to specialized implementation agents (AI Development Units, ADUs) that implement in parallel under the workspace's guardrails.

\subsection{\texorpdfstring{\raisebox{0.1em}{\defEntity{PhaseAColor}{Phase A}}}{Phase A} PLAN -- Workspace Configuration }

The Plan phase establishes guardrails and allows for steering the development context before code is generated.

\iviaparagraph{Idea Brainstorming and Issue Board}
The human developer initiates a project by defining high-level goals and architectural constraints, thereby establishing the foundational context that feeds all AI agents.
Through a brainstorming chat, the \bonsai users explore ideas and tasks collaboratively with an AI assistant; each idea is structurally evaluated for feasibility and complexity, receiving a \emph{go}, \emph{refine}, or \emph{pass} recommendation before promotion to the Issue Board.
Promoted ideas are decomposed into a \emph{parent issue} with numbered \emph{child issues}, each assigned an agent type and linked by explicit dependencies forming a directed acyclic graph, avoiding merge conflicts through proper planning.
The Nexus respects this dependency order during execution, deferring any child whose prerequisites have not yet been completed.

\iviaparagraph{Agent Skill Management} Rather
than relying on a single, omnipotent AI, the workspace allows configuring specialized ADUs.
Here, developers define specific capabilities, assign allowed tools, and set rigid constraints for each agent.
By reducing the action space and tailoring the AI's profile to a specific task (e.g., frontend UI vs.\ backend data processing), the system strategically decreases the risk of context-window ballooning and hallucinations.

\subsection{\texorpdfstring{\raisebox{0.1em}{\defEntity{PhaseBColor}{Phase B}}}{Phase B} DESIGN -- VA Application Composition}
The Design phase is the active, collaborative coding environment where the \bonsai{}'s four-layer architecture is leveraged.

\iviaparagraph{Service Registry}
The Service Registry introduced in \cref{sec:core architecture} is exposed through a dedicated management interface that supports registering external services (in addition to automated discovery), role-based access management, and centralized documentation, including endpoint usage examples and code snippets.

\begin{figure}[t]
    \centering
    \includegraphics[width=1\linewidth]{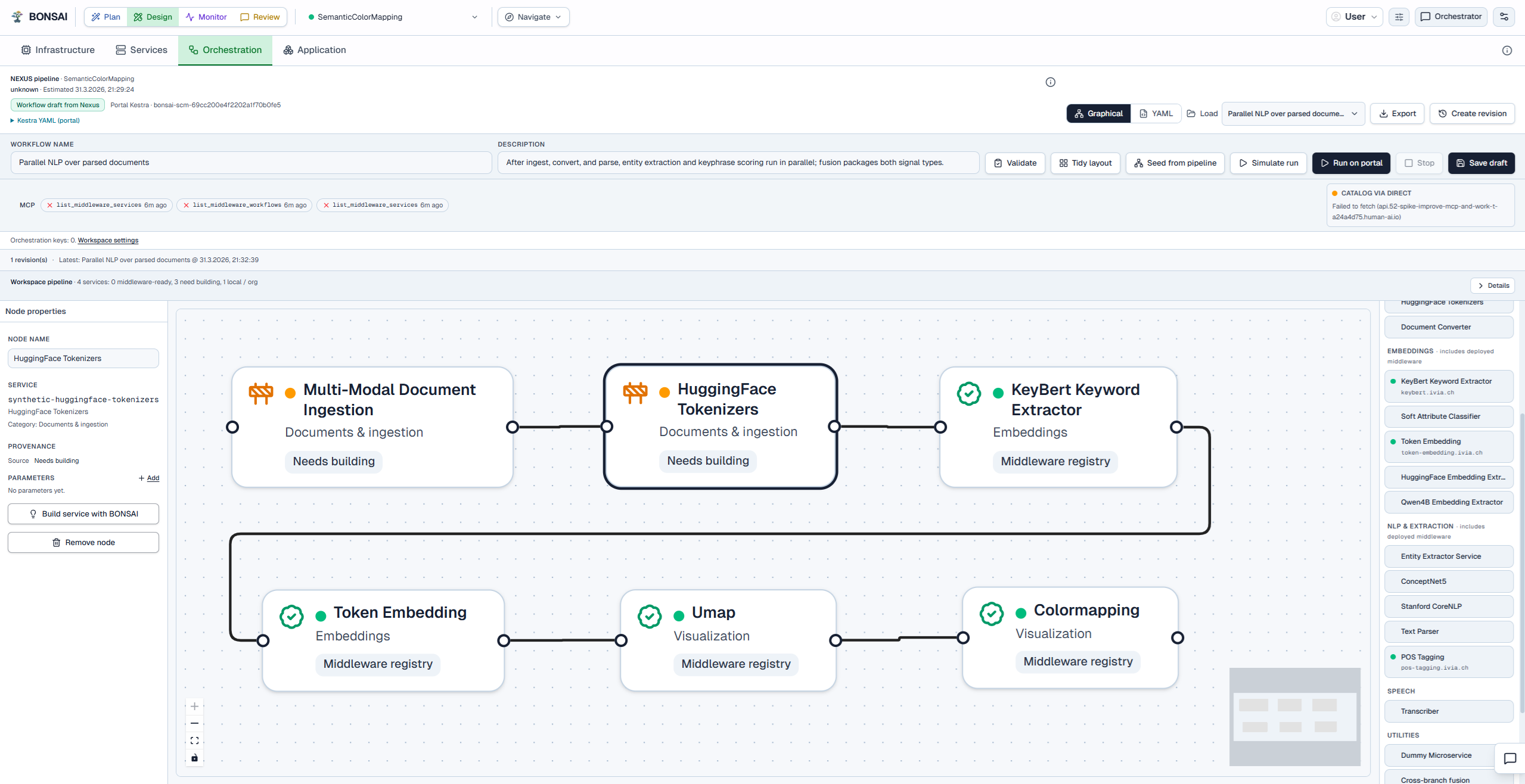}
    \vspace{-1.5em}
    \caption{The orchestration graph shows the constructed Kestra workflow and highlights which services exist or must be built. }
    \label{fig:orchestration}
    \vspace{-1.5em}
\end{figure}

\iviaparagraph{Composing Workflows}
The workspace includes a visual pipeline editor backed by the structural type system (CType) shown in \cref{fig:orchestration}.
Developers select from a toolbox of pre-configured and auto-discovered service nodes and compose them via drag-and-drop in a graph editor.
Valid connections are highlighted in green; type mismatches are flagged in red with a detailed error message on hover, providing immediate design-time feedback before any code executes.
Besides inputs and outputs, nodes can have a special input type called \emph{parameters}; unlike inputs, parameters are not provided by upstream nodes but can be configured as node properties.
Parameters can be set at design time or provided to the flow at runtime.
For example, a node implementing the k-Means algorithm \cite {Lloyd1982LeastSquaresQuantization} might have the parameter $k$ specified at runtime via user input or, if not provided, a plausible value determined by Nexus.

\iviaparagraph{Application View}
Developers write custom code to create the End-User-facing application using their preferred language and frameworks.
An integrated IDE provides direct access to the repository files; for more advanced setups, full-blown desktop IDEs can be used on a local copy of the repository.
This follows the usual Git versioning process, which is also done by the AI agents working on implementation tasks.
From the custom application code, workflows, and service resources can be accessed via simple HTTP requests.
The documentation and examples provided by the Service Registry and the Workflow Compositor provide guidelines and constraints for developers.
Besides the possibility to build a custom VA application, the HTTP-based architecture of the Bonsai Core facilitates an integration into external tools for data processing and visualization (e.g., Jupyter Notebooks\cite{Kluyver2016JupyterNotebooksPublishing} or BI applications\cite{Srivastava2021ReviewStateArt}).

\subsection{\texorpdfstring{\raisebox{0.1em}{\defEntity{PhaseCColor}{Phase C}}}{Phase C} MONITOR -- Supervision \& Provenance}

\begin{figure}[b]
    \centering
    \vspace{-1.5em}
    \includegraphics[width=1\linewidth]{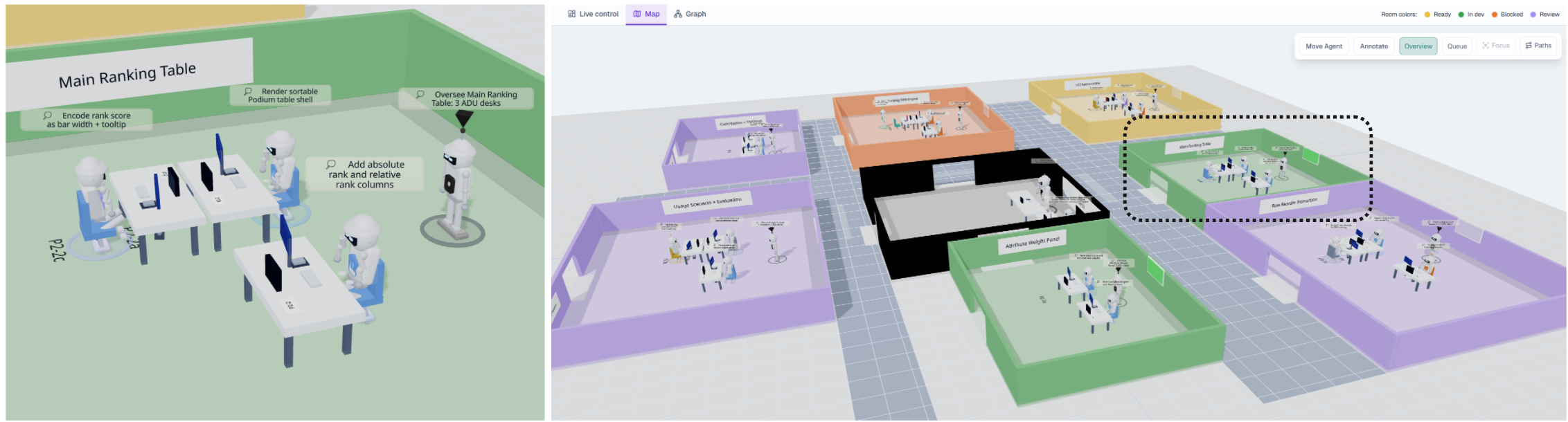}
    \vspace{-1.5em}
    \caption{The Agent Map provides a real-time, spatial overview of current agent activities, statuses, and feature room assignments.}
    \label{fig:agent-map}
\end{figure}

\begin{figure}[t]
    \centering
    \includegraphics[width=\linewidth]{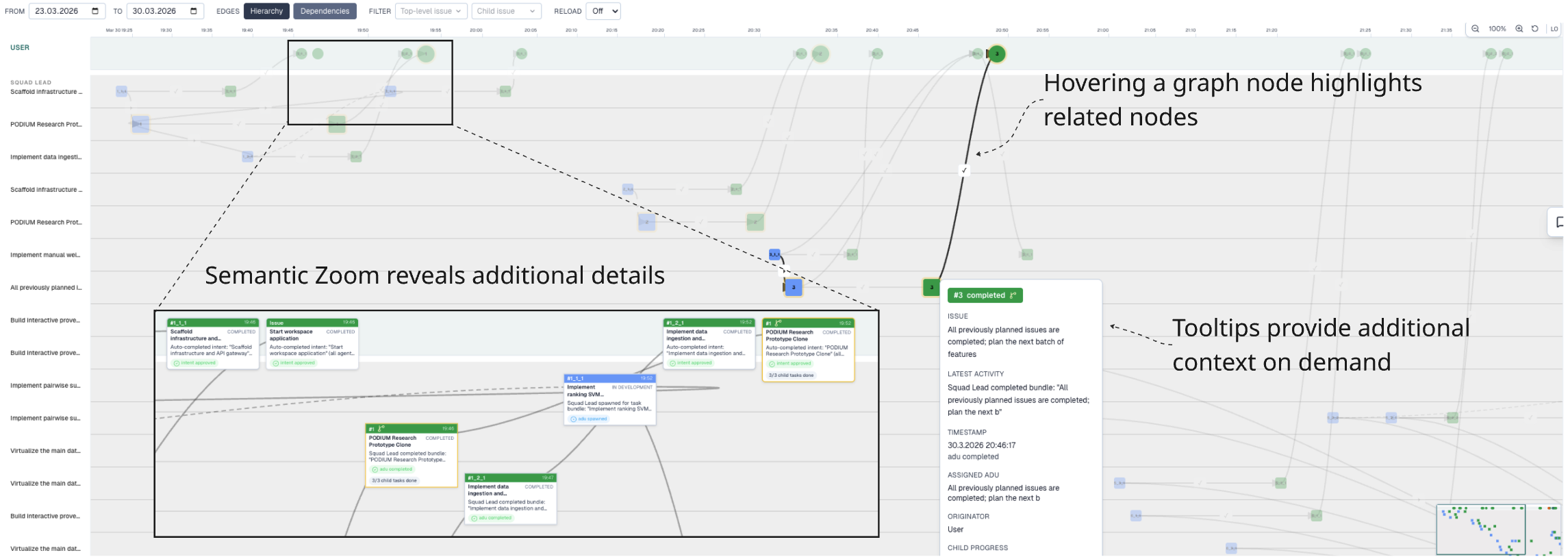}
    \vspace{-1.5em}
    \caption{The provenance visualization shows the development process at different levels of semantic zoom. }
    \label{fig:provenance}
    \vspace{-1.5em}
\end{figure}

The Monitor phase replaces traditional, static logging with dynamic transparency, splitting oversight into two complementary dimensions: the \textit{agent map} for spatial monitoring of ongoing actions and the \textit{history and provenance} component for reviewing past decisions.

\begin{wrapfigure}[6]{r}{0.36\linewidth}
    \vspace{-12pt}
    \begin{minipage}{\linewidth}
        \centering
        \includegraphics[width=\linewidth]{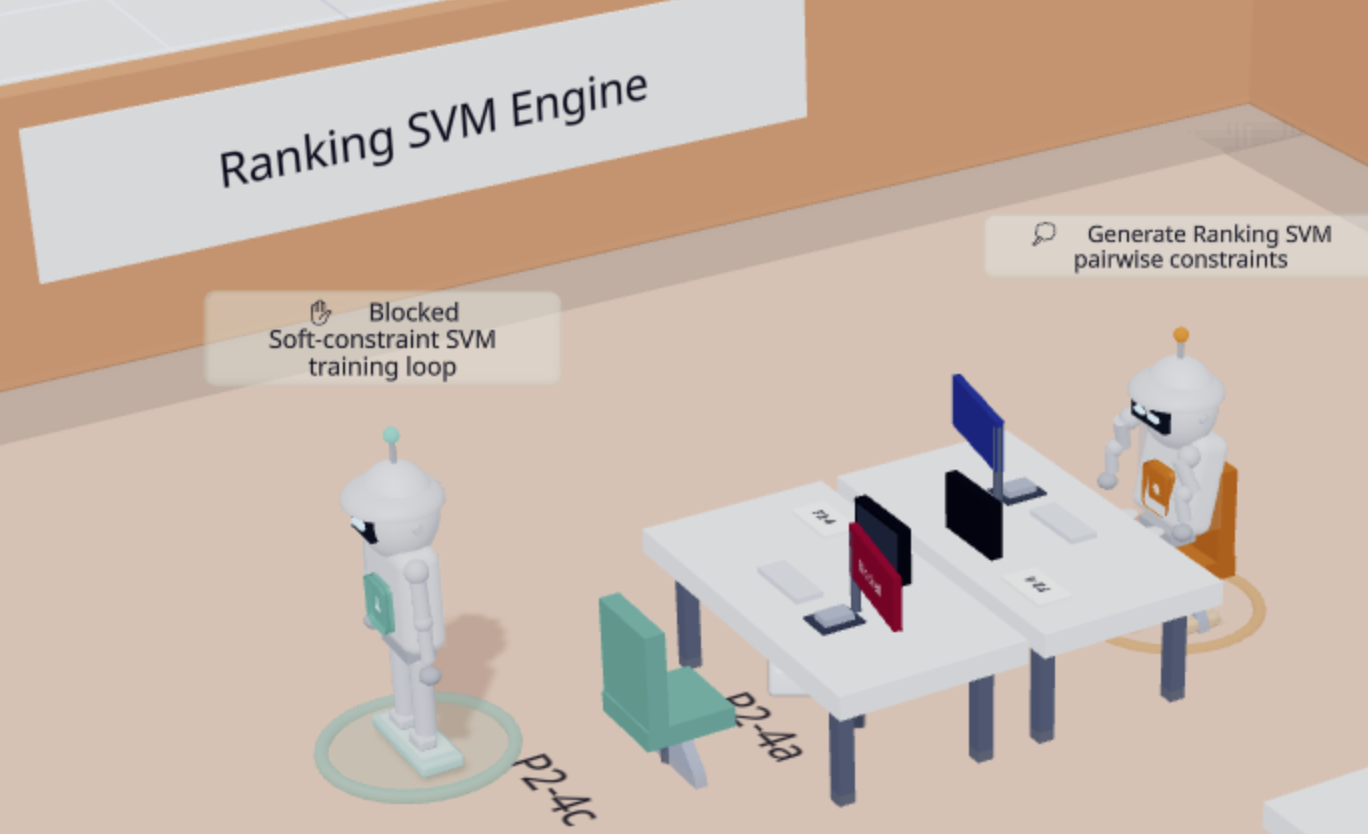}
    \end{minipage}
\end{wrapfigure}
\iviaparagraph{Agent Map View}
\Cref{fig:agent-map} demonstrates how this view provides a real-time overview of all active ADUs, rendered as an interactive 3D scene where each agent is placed within feature rooms corresponding to parent issues on the Issue Board.
Agent \emph{type} is encoded through distinct color and shape, while agent \emph{status} (in development, blocked, queued, in review) is conveyed through visual accents, enabling the developer to assess fleet health at a glance.
There are interactive modals for agents and rooms showing more details about their states, child issues, and merge progress.

Crucially, the Monitor phase is not read-only.
Through the map's integrated \emph{Conductor} control panel, the human can intervene directly: commenting on a running agent to provide additional context, injecting constraints mid-execution, pausing, terminating, or redirecting tasks, and responding to clarifying questions surfaced by blocked ADUs.
This preserves mixed-initiative agency throughout the implementation cycle rather than limiting human input to pre-flight configuration.

\iviaparagraph{History \& Provenance Component}
The provenance component shown in \cref{fig:provenance} captures a fully traceable state history, documenting every architectural decision and treating agency handoffs as first-class provenance events to ensure complete auditability across the co-development process. 
We visualize provenance as a directed acyclic graph (DAG) where the x-axis encodes time and the y-axis organizes actors into horizontal swimlanes:
a dedicated lane for user interactions at the top, followed by one lane per Squad Lead (SL), and finally groups of ADUs (e.g., backend, frontend, database, web design),
each subdivided into rows for individual agent instances.
Lane headers stay pinned during panning to keep actor attribution always visible,
and all actors are consistently color-coded across every view in the workspace.

To keep the potentially large provenance space navigable, we employ a four-level semantic zoom~\cite{bederson1994pad}.
At zoom level~ZL0, only high-level parent issues are shown for a compact project overview.
At~ZL1, every provenance event is rendered as a small, color-coded circle; correction events use an inverted-triangle glyph so that disruptions remain immediately recognizable.
At~ZL2, nodes expand into badge-header cards that show the actor's name, a one-line summary, and a timestamp.
ZL3 augments these cards with pill-shaped links to the corresponding conversation log, git diff, or agent detail view.

Two visually distinct edge types connect provenance nodes: solid edges encode causal relationships (e.g., ``delegated to,'' ``completed''),
while dashed edges denote informational influence (e.g., ``informed,'' ``reviewed'').
Because orchestrator actions frequently cluster within seconds of each other,
a compressed time scale shrinks idle gaps while preserving true proportions within event clusters,
with axis-break marks signaling compressed regions.
A corner minimap provides an overview of the full time range, enabling click-to-navigate interaction.

\subsection{\texorpdfstring{\raisebox{0.1em}{\defEntity{PhaseDColor}{Phase D}}}{Phase D} REVIEW -- Evaluation \& Iteration}
The final phase ensures that the co-developed VA workflow is robust, scalable, and production-ready before deployment.

\iviaparagraph{Merge-Gated Completion and Agency Handling}
Once humans or agents complete development of an issue, it is moved to the Review column of the issue board.
From there, it can advance to Completed only after its feature branch has been merged.
We enforce this constraint at the system level rather than leaving it to the developer or agent discipline.
This strict quality gate closes the human-AI collaboration loop: by default, only human users can trigger a merge from the review queue, and only after a successful merge does the corresponding Issue Board card advance.
To calibrate development speed against control, agency sharing can be calibrated by defining rules on when Nexus may auto-review issues instead of humans.
For parent issues, all child tasks must individually pass this gate before the parent's integration branch is merged into the main branch, ensuring that no partial or unreviewed work is silently introduced into the production codebase.

\iviaparagraph{Iterative Live Application Review and Feedback Loop} Rather than limiting review to code diffs or static screenshots, the workspace embeds the running application in a live preview panel.
Nexus starts the application by scanning for free ports on the host system.
After each merge to the main branch, the development server automatically reloads, allowing users to evaluate the artifact in its intended runtime context.

\begin{figure}[t]
    \centering
    \includegraphics[width=1\linewidth]{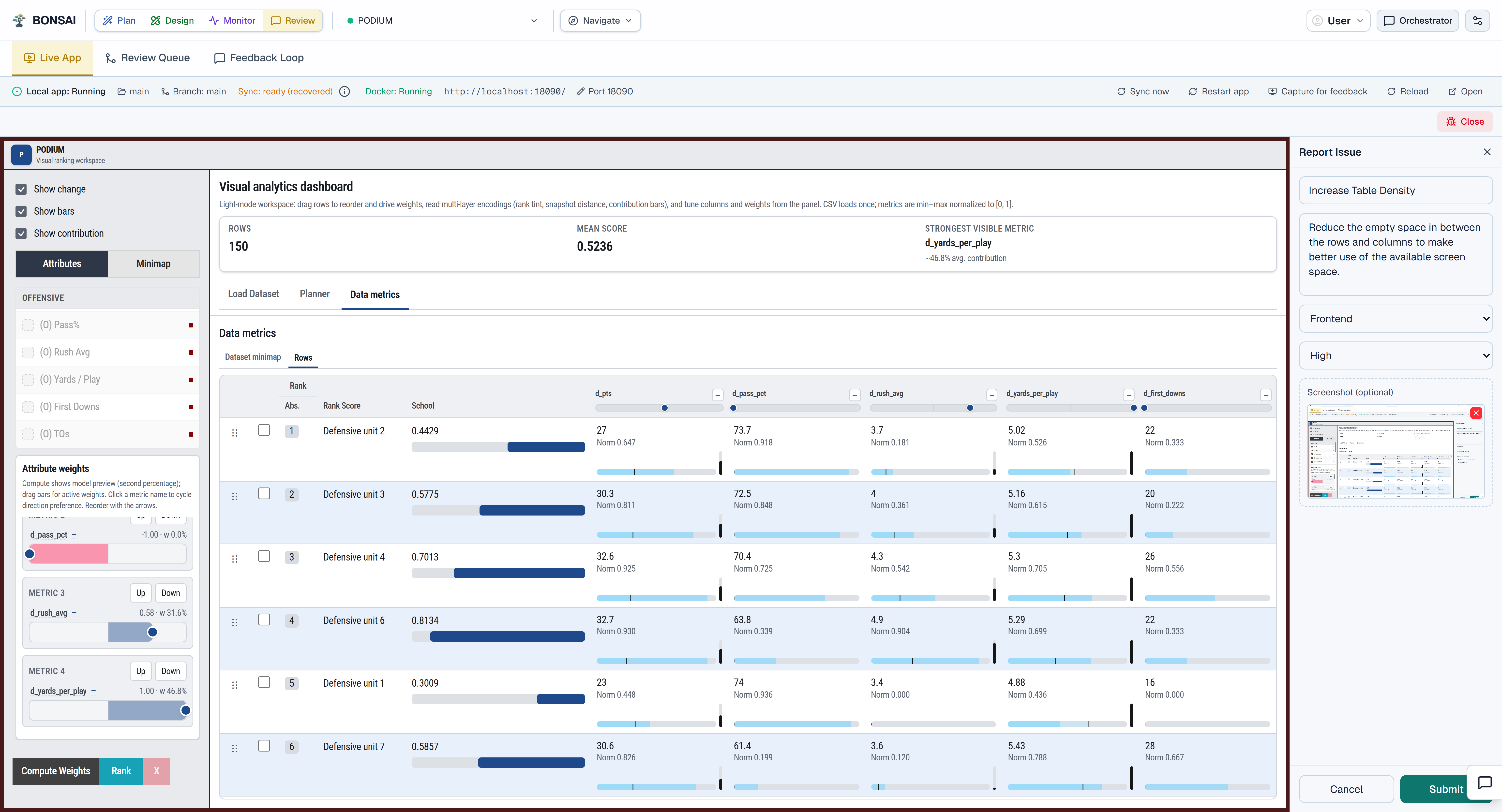}
    \vspace{-1.5em}
    \caption{The \textit{Live App} view shows a running instance of the PODIUM app under development to enable quick testing. Identified bugs and issues can be reported via the sidebar and are added to the issue board. } 
    \label{fig:feedback-loop}
    \vspace{-2em}
\end{figure}

We distinguish two levels of feedback.
At the AI level, the Squad Lead's acceptance-criteria validation can reject worker output and automatically re-queue failing ADUs with diagnostic context, without human intervention.
This also means that branches leading to conflicts that cannot be resolved by the ADUs must either be reimplemented or resolved by human experts.
At the human level, the Review phase embeds a lightweight issue-reporting mechanism directly into the live application view (see \cref{fig:feedback-loop}): while inspecting the running prototype, the developer can directly file new issues from a sidebar, optionally attaching an automatically captured screenshot.
The reported issue is added to the task backlog with status \emph{planning}, feeding back into the Plan phase as a first-class development task, subject to re-planning and re-assignment, similar to the outputs of the brainstorming session.
Both loops converge on the provenance graph, where each re-queue, rejection, or newly filed issue is recorded as a distinct node capturing what was rejected, why, and by whom, ensuring that the graph documents the full trajectory of refinement rather than merely the final state.

\section{Integration of Agentic AI}
\label{sec:integration-of-agentic-ai}

Collaborative VA application development produces high-volume, ambiguous team dialogue.
To translate this naturalistic input into actionable, traceable work while preserving human agency, \bonsai integrates agentic AI as a structured, governed participant.

\subsection{Cognitive Model: Agent Hierarchy}
\label{subsec:cognitive}

We structure agentic participation through three hierarchical roles that mirror a software development team.
The top-level agent, the \emph{Nexus}, monitors the collaborative transcript and parses natural-language dialogue into typed, confidence-scored intents.
It maintains a prioritized backlog, allocates responsibilities across specialized roles, and enforces policy gates that regulate automation throughput.
For multi-concern tasks requiring two or more distinct agent types, the Nexus delegates work packages to \emph{Squad Lead} agents.
The Squad Lead decomposes a parent issue into independent child tasks, extracting explicit acceptance criteria, and enforces a four-phase pipeline (planning, clarification, file declaration, implementation); failing outputs are being re-queued with structured feedback for rework.
Actual coding is performed by \emph{AI Development Units}, specialized sub-agents provisioned on demand from 16 registered types (e.g., frontend, backend, database), each constrained to a single architectural layer.
This hierarchy implements a two-layer cognitive model that explicitly separates reasoning from execution.
The Nexus operates in the \emph{cognitive} layer: structuring messy, exploratory input into concrete decisions, analogous to sensemaking support in visualization research. 
Once intents are structured, they are handed to the \emph{operational} layer, where the Squad Lead and ADUs map decisions onto the formal REST APIs and validated workflow graphs of the layered core architecture. 
The underlying runtime is never LLM-driven as agentic behavior acts strictly as a bridge from informal human intent to formal, auditable pipeline assets.

\begin{figure}[b!]
    \centering
    \vspace{-1.5em}
    \includegraphics[width=1\linewidth]{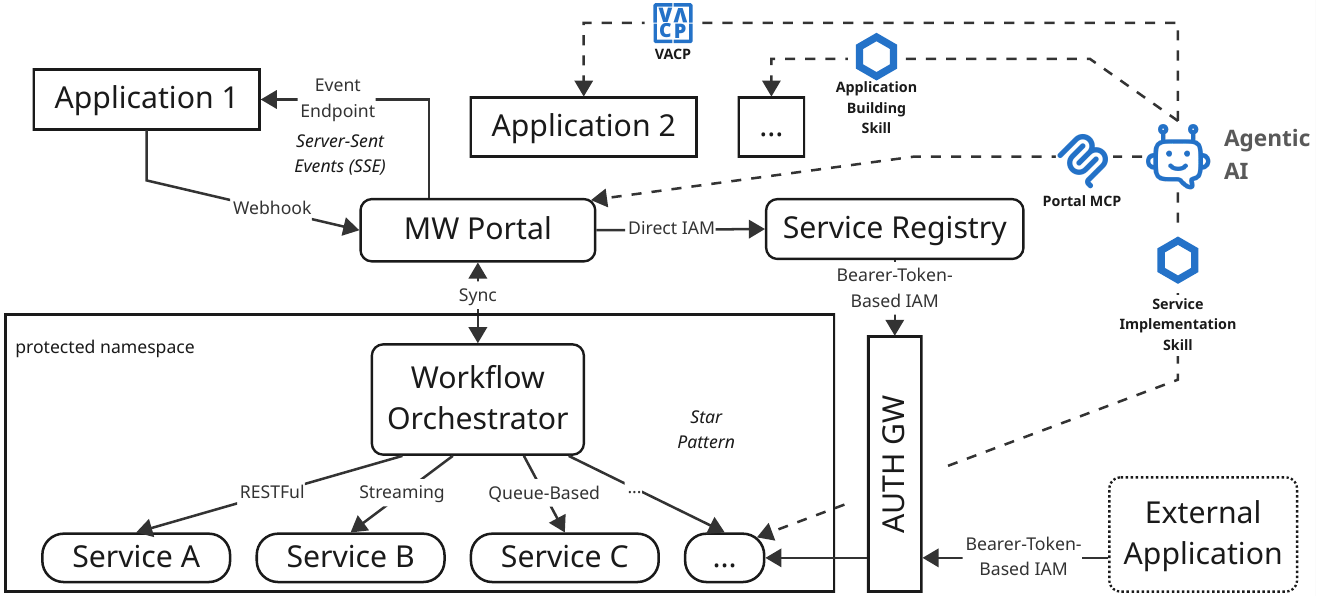}
    \vspace{-1.5em}
    \caption{Agentic integration points within the \bonsai{} architecture.
    AI agents connect through three levels: \emph{Skill Files} for service and application development, the \emph{MCP server} for orchestration-layer operations, and \emph{VACP} for structured interaction with the application's internal state.}
    \label{fig:mwComponents}
\end{figure}

\subsection{Development Workflow}
\label{subsec:workflow}

When a human developer introduces a new directive, \bonsai does not generate code immediately.
Instead, the Nexus classifies the intent (e.g., directive versus exploratory hypothesis) and performs \emph{active service mining}: it queries the L2 service registry and L3 workflow catalog to determine whether existing \bonsai Core services already satisfy the requirements.
Services that match are wired directly into the pipeline.
When no admissible service exists, Nexus spawns SLs and ADUs to implement, containerize, and register a new L2 service.
This new service is subject to the same interface contracts and admission checks that govern manually built components.
This reuse-first strategy ensures that the service catalog grows organically with each development cycle while minimizing redundant reimplementation.

As illustrated in \cref{fig:mwComponents}, \bonsai offers agentic integration at three levels.
At the service level, curated \textbf{Skill Files} encode layer-specific coding conventions, architectural patterns, and domain guardrails; they equip each ADU with precisely the capabilities and context knowledge needed to build or extend a component within its assigned task.
At the orchestration level, a dedicated \textbf{MCP Server} exposes design-time operations (service discovery, workflow validation, DAG composition, and deployment) as structured, bounded function calls.
Because these MCP tools mirror the same operations available through the graphical workflow designer, the identical guardrails governing human composition also govern AI-generated workflows, with no privileged code path for automated agents.
In theory, human developers could even create and register new services outside of \bonsai, if needed.
ADUs are aware of the Kestra workflow specification~\cite{kestra2024} through their skill files and can compose and register orchestration pipelines programmatically via the MCP Portal.
At the application level, the \textbf{Visual Analytics Context Protocol (VACP)}\cite{Staehle2026VacpVisualAnalytics} embeds an MCP server directly into the application under development, granting agents structured access to internal application state and data; this targeted integration enables domain-specific VA tasks (e.g., coordinated view updates, parameter steering) that pure browser-based agent control cannot reliably achieve.

ADUs are scheduled as external cloud agents that receive a strictly scoped task package including an encapsulated sandbox, skill files, file-lock boundaries, and a dedicated Git feature branch.
They submit code artifacts, without access to orchestration or to each other's working context, preserving the bounded execution model that operationalizes~\refEntity{R2}.

\subsection{Execution Constraints and Governance}
\label{subsec:governance}

To ensure that the development workflow described above remains a governed process rather than unconstrained parallelism, we enforce several complementary mechanisms.

\iviaparagraph{Policy Gates}
Before any ADU is spawned, the Nexus evaluates a series of preconditions: a configurable confidence and predefined timeout thresholds determine whether intents are auto-approved or (temporarily) held for human confirmation; concurrency caps limit simultaneous agent execution; dependency resolution defers tasks whose prerequisites remain incomplete; and file-lock checks prevent agents from writing to the same files.
A configurable briefing pause between spawn and execution gives human developers a window to inspect or abort planned agentic tasks before code generation begins.

\iviaparagraph{Coordinated Staging}
For multi-concern initiatives, the SL delegates orthogonal child tasks, minimizing overlapping files across concurrent ADUs.
This also means that a single ADU can perform multiple tasks in a single run when similar files must be modified.
When all ADU workers are finished, the Squad Lead validates results against the required acceptance criteria using a structured evaluation; failing outputs are rejected with specific feedback and re-queued, up to a configurable number of cycles before escalation to human review. 

\iviaparagraph{Branch Isolation and Merge Governance}
Every ADU operates on a dedicated, separate Git feature branch.
Parent issues receive an integration branch; child ADUs merge back into it upon successful completion.
Only after the combined results of all ADUs pass the Squad Lead's review, Nexus permits merging the integration branch back into the main branch.
When outstanding branches exceed a configurable threshold or merge conflicts arise, a specialized Merge ADU handles conflict resolution, ensuring that the merge process itself remains governed.
If merge conflicts cannot be resolved, the results of corresponding branches are summarized and retriggered for implementation.

\iviaparagraph{Architectural Drift Prevention}
Two mechanisms jointly address~\refEntity{R2} by counteracting context degradation.
First, the Nexus's service mining actively surfaces reuse opportunities before any new code is written, keeping human developers informed of what the AI reuses versus what it generates from scratch.
Second, skill files do not merely provide coding guidance; they encode the structural conventions each service must satisfy (e.g., OpenAPI contract requirements, admission criteria, layer-specific interface patterns).
Together, service mining and skill-encoded guardrails ensure that AI-generated components conform to the platform's architectural standards rather than silently introducing drift.

\section{System Validation \& Use Cases}
\label{sec:evaluation}

Evaluating a comprehensive architectural framework for human-AI co-development presents a methodological challenge that time-boxed controlled studies cannot adequately address: the longitudinal friction of dependency management, iterative design, and multi-agent coordination in real VA projects unfolds over weeks or months, not a single session.
We therefore adopt a systems-validation methodology common in HCI and visualization infrastructure research and evaluate the \bonsai workspace through two representative replication use cases~\cite{cutler2026revisit}.

Rather than cataloging outcomes in abstract capability buckets, we present each case as a walkthrough tracing workspace views, human decisions, and agent handoffs. 
The \textit{Semantic Color Mapping}~\cite{elassady2022semantic} case tests how effectively the workspace leverages an existing L2-service-catalog to simplify the build scope through reuse.
The \textit{PODIUM}~\cite{wall2018podium} case tests the converse: zero catalog matches, forcing the architecture to decompose a paper into independent services built from scratch.
Together, the two cases span the reuse spectrum and reveal how the same orchestration framework supports both extremes.

\subsection{UC1: Component Extraction: Semantic Color Mapping}
\label{subsec:uc1}

The \textit{Semantic Color Mapping} (SCM) pipeline~\cite{elassady2022semantic} is a recurring building block across several VA systems, comprising a staged process from aggregated text data through vector representations to perceptual color assignment.
In practice, it has historically been embedded deep within monolithic codebases such as the \textit{generAItor} system~\cite{Spinner2024GeneraitorTreeLoop}.
We evaluate whether \bonsai{} can extract this logic from the existing monolith and reconstitute it as a composable workflow built entirely from existing middleware services, avoiding redundant reimplementation.

\begin{figure}
    \centering
    \includegraphics[width=\linewidth]{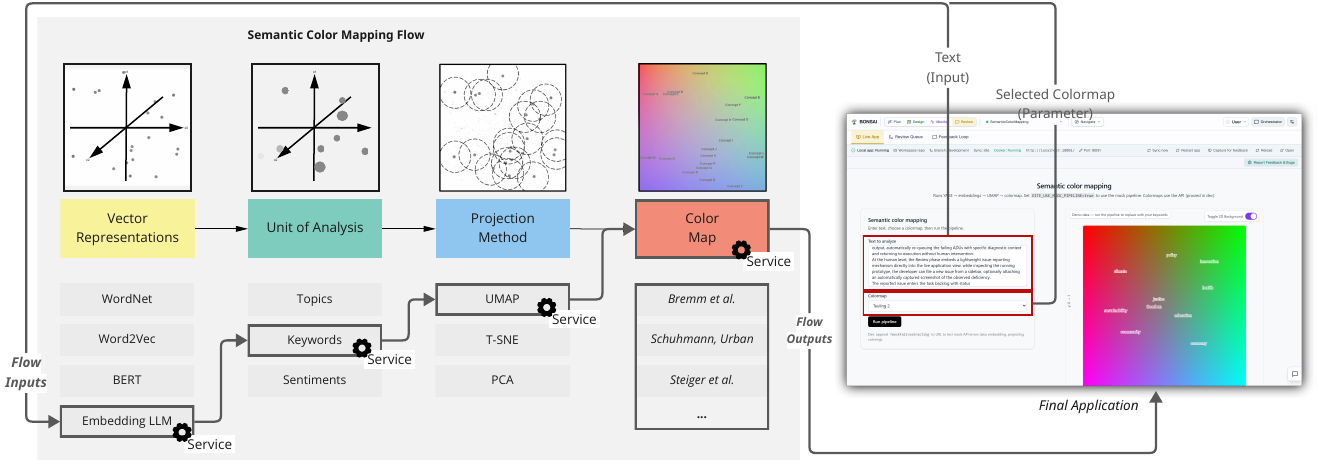}
    \vspace{-1.5em}
    \caption{The SCM pipeline~\cite{elassady2022semantic} comprises multiple steps with clear interfaces before visual encoding is possible. The separate nodes in this pipeline represent separate building blocks covered by independent services provided by the \bonsai Core.}
    \label{fig:semantic-color-mapping}
    \vspace{-2em}
\end{figure}

\iviaparagraph{Development Walkthrough}
Cooperating with Nexus, we began in \refEntity{Phase A} by decomposing the SCM process into its constituent stages: keyword extraction, embedding computation, dimensionality reduction, and 2D color-map assignment.
We also clearly defined how each of these steps must be sequentially connected for seamless transitions.
Moving on to \refEntity{Phase B}, we asked Nexus to assemble a workflow that constitutes the SCM pipeline.
Leveraging the MCP tool for service catalog search, \bonsai{} identified that four of the required processing stages were available as microservices, ready for direct use.
Consequently, the agent determined that mostly orchestration-level plumbing was required to connect the existing services into a valid DAG.
During the first composition attempt, the CType system flagged a schema mismatch between the embedding service's output format and the projection service's expected input structure.
Because the error message localized the incompatibility to a specific field-level type conflict, the ADU resolved the discrepancy by inserting a lightweight adapter node, addressing the input and output mismatches.
Throughout both attempts, we tracked the ADU's status transitions on the agent map in \refEntity{Phase C}.
In the second iteration, the workflow passed all structural validation checks for the \bonsai Core and executed end-to-end, producing semantically meaningful color mappings consistent with the original description.
This yielded a working mini-application that we inspected in \refEntity{Phase D}, exposing input forms and a visualization of the workflow's output.
\Cref{fig:semantic-color-mapping} shows the interface, with arrows indicating the data flow between the application and \bonsai's Core.

\iviaparagraph{Lessons Learned}
This case study surfaces three observations relevant to the design of mixed-initiative co-development workspaces.
First,  \emph{without explicit instruction, Nexus distills the requirements for the given situation} to identify whether there are existing service modules that can already be employed instead of reinventing the wheel.
Early versions of \bonsai{} lacked this behavior; dedicated skill files that prioritize catalog search over regeneration fixed it.
Second, \textit{the structural type system is essential for third-party service reuse}.
When an ADU implements a service itself, it implicitly controls both sides of every interface.
Reusing a pre-existing service removes this control; the ADU must conform to an interface it did not author.
The CType system closes this gap by disclosing precise, field-level incompatibilities at design time, enabling the ADU to correct mismatches before runtime execution.
Third, \textit{reusability dramatically reduces implementation effort}.
Because the middleware already contained all required processing stages, validated, and production-grade, the entire extraction task was reduced to the orchestration layer.
Compared to a from-scratch reimplementation, this approach required only two ADU iterations.
It demonstrated that a well-populated service catalog combined with strong interface contracts can transform complex feature migration into a lightweight wiring exercise.

\subsection{UC2: System Reconstruction: PODIUM}
\label{subsec:uc2}

Wall et al.'s \textit{PODIUM}~\cite{wall2018podium} couples a tabular ranking interface with a Ranking SVM trained from drag-and-drop preferences.
We selected it to contrast the previous Semantic Color Mapping (SCM) use case: service mining found no admissible L2 matches in the registry, meaning the architecture had to decompose the paper into independent services built entirely from scratch and composed through REST interfaces.
A further objective was to demonstrate that \bonsai{}, with human steering, can replicate a complex paper from a single source, yielding a high-fidelity, interactive prototype for pedagogical or demo use.
Rather than relying on unconstrained ``vibe-coding,'' which typically yields a brittle, opaque monolith, we used \bonsai{} to produce a rigorously architected, modular system that students can dismantle, inspect, and build upon.

\begin{figure}
    \centering
    \includegraphics[width=1\linewidth]{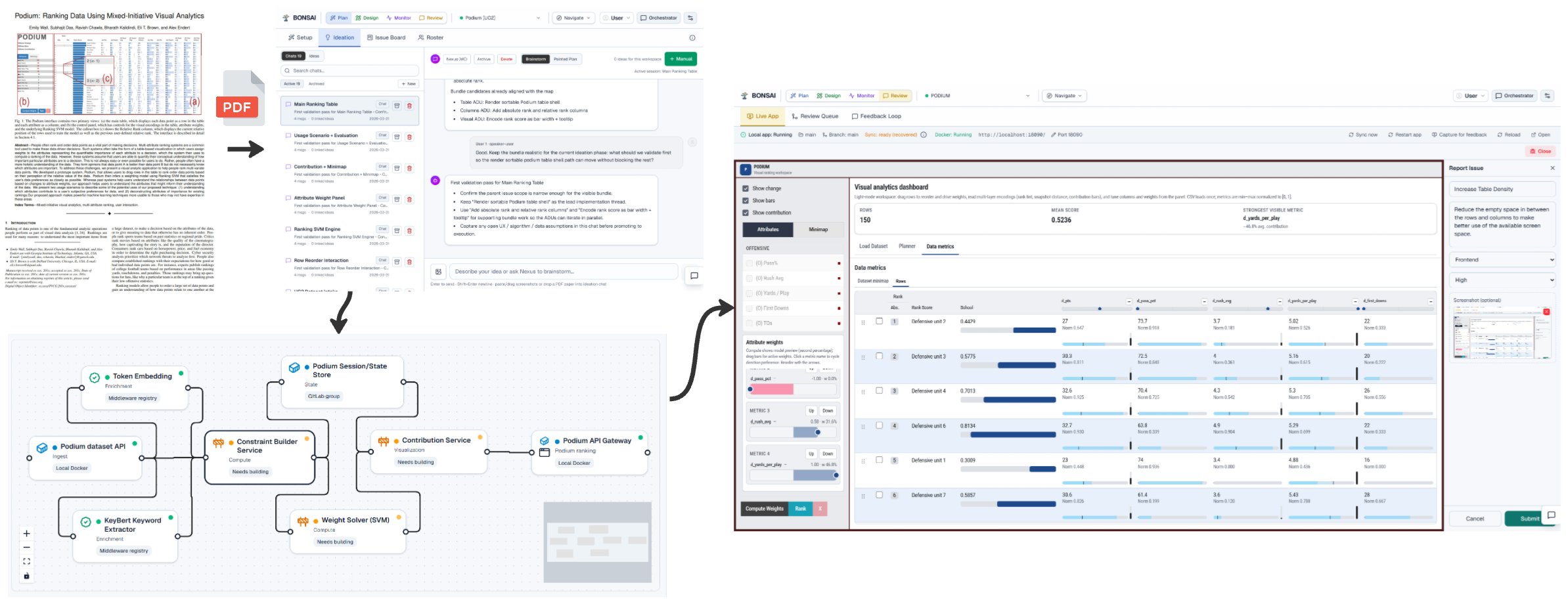}
    \vspace{-1.5em}
    \caption{Given a publication (as PDF), users can employ the brainstorming chat to extract relevant parts of existing techniques, identify service, workflow, and implementation requirements, reuse existing \bonsai Core services, and reimplement missing modules based on the four stages.}
    \label{fig:podium}
    \vspace{-2em}
\end{figure}

\iviaparagraph{Development Walkthrough}
The whole development process is summarized in~\cref{fig:podium}:
During \refEntity{Phase A}, we initiated the workflow by providing Nexus with a direct link to the \textit{PODIUM} manuscript PDF in the brainstorming chat.
Nexus retrieved the document and successfully extracted the core system components.
It then proactively asked clarifying questions regarding the underlying machine learning architecture.
We explicitly directed the Nexus to strictly adhere to the paper's original Ranking SVM rather than introducing modern LLM-based explanation features.
A notable success during this phase was the underlying model's multimodal reasoning: the Nexus was able to process figures from the paper and automatically attach relevant UI mockups and architectural diagrams directly to the child issues.
Subsequently, Nexus delegated the remaining implementation tasks to squad leads via issues on the issue board.
In \refEntity{Phase B}, the squad leads then instructed ADUs to implement the required backend services as well as the visualization frontend.
During \refEntity{Phase C}, we monitored squad-lead delegation and ADU progress on the provenance graph, intervening via the Conductor panel when agents became blocked.
After an initial implementation had been produced, we performed several feedback loops via the Live App view's annotation feature (see \cref{fig:feedback-loop}) to adjust the visual styling and interaction affordances in \refEntity{Phase D}.
Ultimately, we received a set of microservices and a modern frontend that could serve as a showcase for students to demonstrate the service- and worflow-based development process.

\iviaparagraph{Lessons Learned}
This case study surfaces four critical observations regarding the reconstruction of complex academic systems.
First, \textit{agentic paper-to-code reconstruction is a highly viable pathway for reviving legacy research}.
The workspace successfully translated a static PDF description into a live, interactive VA application without requiring access to the original source code.
This demonstrates significant promise for democratizing access to older, unpublished, or deprecated visualization techniques that would otherwise be lost to software rot.
Second, \textit{managing high-complexity deployments requires strict layer separation}.
Unlike simpler extractions, the PODIUM reconstruction involved a frontend, a backend, and seven distinct, interconnected microservices (including a dataset provider and an SVM-based weight solver).
By enforcing the 4-layer architecture, \bonsai ensured that the Application ADUs and Service ADUs could generate code concurrently without tangling the complex state logic.
Third, \textit{strict service registry admission introduces friction during rapid prototyping cycles}.
Because \bonsai requires human review before a newly generated service can be officially admitted and scheduled by the orchestrator, fast development loops can become bottlenecked.
During the PODIUM build, pending components had to be evaluated in a local development environment.
To mitigate this scheduling delay and keep the orchestration workflow intact, we found that developers can temporarily mock service functionality using inline code snippets that Kestra can process until the formal review is complete.

\section{Discussion, Limitations, \& Future Work}

While \bonsai{} demonstrates how bounded orchestration resolves the
sociotechnical friction of agentic VA development, its design involves
deliberate trade-offs worth examining.
This section reflects on key
design decisions, discusses agency distribution and ecosystem dynamics,
and outlines pathways for future contributions.

\iviaparagraph{Learning Curve of the Layered Architecture}
Adopting \bonsai{} requires developers to internalize the four-layer mental model and interface requirements before contributing a service, which may deter those accustomed to flatter architectures, where immediate code execution takes priority.
To mitigate this, we propose two complementary mechanisms.
First, layer-specific skill files enable developers to engage in a natural dialog with agentic AI about architectural conventions, effectively turning static documentation into an interactive learning resource.
Second, we are developing \textit{Onboarding ADUs} that auto-generate boilerplate OpenAPI contracts and scaffold compliant project structures.
Together, these mechanisms aim to reduce the time-to-first-service while preserving the rigor that underpins cross-project reusability.

\iviaparagraph{Automated Architectural Auditing and Governance}
A practical bottleneck in scaling \bonsai{}'s ecosystem is the manual review required to admit new services, which constrains throughput as the catalog grows.
We see complementary directions for addressing this.
First, \textit{Auditor ADUs} can automate static analysis and compliance checks, flagging non-compliant components before human review.
Second, a \textit{staging mechanism with time-limited deployment} would allow newly developed services to be tested in a sandbox before formal admission.

\iviaparagraph{The Trade-off: Guardrails vs. Creative Flexibility}
The core philosophy of \bonsai relies on the premise that unconstrained ``vibe coding'' must be deliberately pruned using explicit interface contracts.
However, this introduces a mandatory, contract-first development paradigm.
In traditional, monolithic environments, developers can fluidly and chaotically experiment with UI and backend logic simultaneously. \bonsai{}'s strict separation of concerns requires the Nexus and human developers to define data schemas and OpenAPI specifications \textit{before} the Application ADUs can effectively generate the interface.
While this upfront sociotechnical friction drastically reduces downstream debugging and architectural drift, it inherently limits the unstructured, ``blank canvas'' momentum that some developers prefer during early-stage exploratory ideation.

\iviaparagraph{Agency Distribution Across the Human-AI Spectrum}
\bonsai{}'s four-phase workflow makes no assumption about agency distribution between human and AI contributors.
At one extreme, a developer completes all phases manually, treating the workspace as an architectural scaffold.
At the other end, the Nexus decomposes, delegates, and validates entire feature sets while the human acts solely as a conductor.
In practice, a natural middle ground emerges: developers retain control over design-critical decisions (goal setting, interface contracts, acceptance criteria) while delegating repetitive tasks to ADUs.

This flexibility mirrors the ongoing shift in software engineering, where developer roles evolve from writing code toward orchestrating and steering AI-generated artifacts.
\bonsai{} accommodates this transition by design: the same guardrails, provenance tracking, and review mechanisms apply regardless of whether a human or AI authored a component.
As frontier models improve, we expect the automation boundary to shift further toward high-level design tasks, making \bonsai{}'s bounded orchestration model increasingly relevant.

\iviaparagraph{Ecosystem Dynamics \& Reuse-First Development}
The two case studies reveal a self-reinforcing dynamic at the core of \bonsai{}'s service ecosystem.
\hyperref[subsec:uc2]{UC2} required building all services from scratch, yet the resulting component became available for future projects upon admission.
\hyperref[subsec:uc1]{UC1} demonstrates the benefit: most processing stages already existed as microservices from other projects, mostly reducing pipeline construction to orchestration-level wiring in only two ADU iterations.

This flywheel effect is a direct consequence of \bonsai{}'s architectural enforcement of modularity and reuse.
Because every service must satisfy typed interface contracts and pass admission checks independently of any particular application, components produced by one project are natively composable in contexts their original developers never anticipated.
Additionally, MCP-exposed service discovery ensures agents receive only relevant service definitions for their current task, preventing context-window saturation as the catalog grows.

\section{Conclusion}

The rapid evolution of generative AI offers unprecedented prototyping speed but also introduces severe sociotechnical risks, often yielding opaque, tightly coupled monoliths in which developers lose the causal thread of system evolution.
To bridge the gap between unconstrained coding and rigorous software engineering, we introduced \bonsai, a framework that replaces unconstrained full-stack code generation with bounded orchestration across a four-layer architecture.
Through explicit interface contracts, \bonsai distills the context window for specialized ADUs,  ensuring analytical logic is cleanly extracted into reusable microservices.
Crucially, its structured four-phase workspace treats semantic provenance as a first-class citizen, guaranteeing a legible history of agency handoffs.
Ultimately, much like cultivating its namesake, \bonsai demonstrates that sustainable multi-agent development requires deliberate structure and guided oversight, ensuring that next-generation Visual Analytics applications remain scalable, maintainable, and firmly under human expert control.
The system and its source code will be available at \texttt{\href{http://bonsai.ivia.ch}{bonsai.ivia.ch}}.

\clearpage
\bibliographystyle{abbrv-doi-hyperref}
\bibliography{template}

\end{document}